\documentclass{article}
\usepackage[utf8]{inputenc}
\usepackage[margin=1.25in]{geometry}
\usepackage{setspace}
\usepackage{algorithmic}
\usepackage{amsmath,amssymb,amsfonts}
\usepackage{algorithmic}
\usepackage{graphicx}
\usepackage{textcomp}
\usepackage{bm}
\usepackage{gensymb}
\usepackage{amsfonts}
\usepackage{textcomp}
\usepackage{wrapfig,colortbl}
\usepackage[style=nejm, 
citestyle=numeric-comp,
sorting=none,
backend=bibtex]{biblatex}
\title{Towards 3D fully randomized frequency-domain reconstruction of the speed of sound in breast ultrasound computed tomography}
\author{Luca A. Forte
\thanks{Faculty of Science, Engineering and Business, Division of Engineering, University of Dundee, Nethergate, Dundee, DD1 4HN, Scotland, UK (e-mail:{lforte001@dundee.ac.uk).}}}

\bigskip

\date{\today}
\begin{document}
\maketitle

\begin{abstract}
Ultrasound computed tomography is emerging as a promising diagnostic imaging tool. 2D geometries suffer from notorious out-of-plane scattering artifacts. Image reconstruction can be achieved with frequency-domain full waveform inversion and it can be further accelerated by randomly phase-encoding the elementary sources. In this manuscript, we extend our previous results for the 2D geometry of a ring-array to the 3D cylindrical geometry of multiple rings. In particular, we consider the cases of an elementary source described by a single array element (quasi-omni-directional transmission), a line source and a focused transmission respectively. With differences in image quality, we prove that a fully randomized frequency-domain inversion in 3D is capable to reconstruct portions of a human breast surrounded by the cylindrical geometry and detect mm-size masses of varying contrast in dense breast, in reasonable computing times, thus opening the concrete possibility to the design of 3D imaging devices with high sensitivity levels. The methods are applicable to multiple tomographic geometries in 3D and, in principle, can be integrated into next generation medical ultrasound scanners.
\end{abstract}

\section{Introduction}
		
		The framework of ultrasound computed tomography is emerging as a promising diagnostic imaging tool for clinical applications like breast imaging, \cite{DuricLast}, \cite{Wiskin2}, potentially competing with image quality of a MRI scan (the latter often with contrast injection); clinical applications to body parts with high contrast also show promising results \cite{Wiskin3}-\cite{Wiskin4}. It is commonly agreed that 2D systems are affected by out-of-plane scattering, and that 3D systems may solve the associated limitations. This has been shown convincingly in the system described in \cite{Wiskin2}: a fast 3D frequency-domain reconstruction is possible by solving the Helmholtz equation within the so-called paraxial approximation, the latter being enabled by the planar geometry of the system. This approach cannot be translated to other tomographic geometries. Recent attempts to recover the speed of sound with the standard formulation of FD-FWI in 3D have been made in \cite{Scherzer} for a rectangular geometry of an imaging box (224 omni-directional sources, omni-directional array elements in receiving) and in \cite{Duric3D} for a cylindrical geometry (32 rings, 256 line sources in transmission, 256 $\times$ 32 omni-directional array elements in receiving). In the first, the Helmholtz equation is solved with the LU approach: this requires 800 GB of memory on the CPU and it only allows to invert frequencies up to 300 kHz on a voxel of size $\sim$1 mm (60 iterations in total, 5 hours). The in-feasibility of the LU approach in 3D is a known recurring issue, and it's unlikely that it will have any role for 3D ultrasound tomography. In the second, the Helmholtz equation is solved with a modified Born series approach \cite{Ivo}, inverting from 200 kHz to 800 kHz (50 kHz step, 5 iterations per frequency) on an isotropic voxel of 0.4 mm (elapsed times are not reported). Memory requirements for Born-series approaches or recent iterative solvers of the Helmholtz equation \cite{Vuik} are ideal, but computing times can be high due to the inherent iterative features of these solvers. Born-series approaches have the advantage that, in theory, they are unconditionally stable, so higher frequencies can be inverted on coarse grids; on the other side, important anatomical features may require finer voxels. As the iterative nature of these solvers cannot be escaped, the computing times can only be beaten down by considering few sources: drastically reducing the number of tomographic views, however, significantly compromises image quality. A solution is to employ source-encoding approaches, that capture all the sources/receivers combinations with an all at once approach. In this manuscript, we extend our recent results \cite{ForteIEEE2026} to a 3D geometry in transmission and in receiving. In particular, we consider three different elementary sources (quasi-omni-directional, line-source and focused transmission) and we show that the stochastic phase-encoding approach is equally robust, with, however, significant difference in the computational gains with respect to the standard FD-FWI formulation. The case of a quasi-omni-directional transmission seems to provide the best image quality. 
The structure of the manuscript is the following. In section \ref{fd_fwi}, we specialize the stochastic phase encoding approach to a cylindrical geometry, by differentiating among three different elementary sources, which is an important novel contribution. In Section \ref{Results}, we present the results of the numerical studies and show that a portion of a human breast can be accurately reconstructed, allowing for the detection of mm-size masses of varying acoustic contrast in dense breast. In the Conclusions section, we summarize our findings and briefly comments on the feasibility of the methods and their adoption by current and future medical ultrasound imaging devices.

\section{Frequency-domain full-waveform inversion} \label{fd_fwi}
In the section, we briefly review the standard approach to FD-FWI and the formulation with stochastic phase encoding. We refer the readers to \cite{DuricLast}, \cite{ForteIEEE2026} and references therein for additional technical details. 

	\subsection{Conventional approach}
In a model-based image formation approach, one usually starts with defining a cost function. In FD-FWI, the single-frequency cost function reads as:
\begin{equation} \label{Cost}
	 \begin{split}
	F(\omega_{_{0}}; \boldsymbol{c}) = \sum_{s \,=\,1}^{N_{_{s}}} \,F^{^{(s)}}(\omega_{_{0}};\boldsymbol{c})  = \\
 \sum_{s \,=\,1}^{N_{_{s}}} \left\Vert\boldsymbol{d}_{s}^{^{(syn)}}(\omega_{_{0}};\boldsymbol{c}) -  \boldsymbol{d}_{s}^{^{(obs)}}(\omega_{_{0}})\right\Vert_{_{2}}^{^{2}} = \\ \sum_{s\,=\,1}^{N_{s}}\,  \left[ \sum_{j^{^{(s)}}}^{N_{_{Rx}}^{^{(s)}}}\left| d_{s,\, j}^{^{(syn)}}(\omega_{_{0}}; \boldsymbol{c})-d_{s,\,j}^{^{(obs)}}(\omega_{_{0}})\right|^{^{2}}\right]
		 \end{split}
\end{equation}
where $s$ labels the individual source (among $N_{s}$ independent sources) and $j^{^{(s)}}$ is an index running along the corresponding receivers $N_{_{Rx}}^{^{(s)}}$. Different sources may have different sets of physical receivers. The vector $\boldsymbol{d}_{s}^{^{(obs)}}(\omega_{_{0}})$ represents the measured data as complex frequency samples; the vector $\boldsymbol{d}_{s}^{^{(syn)}}(\omega_{_{0}};\boldsymbol{c})$ represents the synthetic data as complex frequency samples, obtained by solving the Helmholtz equation, or an approximation. The complex numbers $d_{s,\,j}^{^{(obs)}}(\omega_{_{0}})$ and $d_{s,\, j}^{^{(syn)}}(\omega_{_{0}}; \boldsymbol{c})$ represent the single-frequency samples at receiver $j$ when source $s$ is active (observed and synthetic respectively) . The calculation of the synthetic data requires the knowledge of the speed of sound. The latter may be updated iteratively through the equation
	\begin{equation}
	\boldsymbol{c}_{_{k+1}} = \boldsymbol{c}_{_{k}} -  \alpha_{_{k}}\,\boldsymbol{\nabla}_{_{\boldsymbol{c}}}(\omega_{_{0}},\boldsymbol{x};\boldsymbol{c}_{_{k}})
	\end{equation}
where the step-size may be tuned with line search. The gradient with respect to the speed of sound is simply the sum over all the individual source gradients
	\begin{equation} \label{sumGrad}
	\boldsymbol{\nabla}_{_{\boldsymbol{c}}}(\omega_{_{0}},\boldsymbol{x};\boldsymbol{c}_{_{k}})\, = \mathfrak{Re}\left[\, \sum_{s\,=\, 1}^{N_{_{s}}} \boldsymbol{\nabla}^{^{s}}_{\boldsymbol{c}}(\omega_{_{0}},\boldsymbol{x};\boldsymbol{c}_{_{k}})\,  \right]
	\end{equation}
	
The previous cost function is very general: in particular, it is valid for any configuration of sources and receivers, and it is equally valid in 2D and in 3D. In the following, we refer to an individual source as to an \textit{elementary} source.
		\subsection{Source Encoding}
With source encoding, groups of individual sources are \textit{incoherently} grouped together in a so-called super-source (or super-shot). An iteration-dependent and super-source-dependent random phase is applied to each \textit{elementary} source making the super-source; the same random phase is applied to the corresponding data-set. The encoded cost function reads as:
\begin{equation}
							\begin{split}
	F^{^{(enc)}}(\omega_{_{0}};\boldsymbol{c}) = \sum_{ss \,=\,1}^{N_{_{ss}}}\,F^{^{(ss)}}(\omega_{_{0}};\boldsymbol{c}) = \\
	 \sum_{ss \,=\,1}^{N_{_{ss}}} \left\Vert\boldsymbol{D}_{ss}^{^{(syn)}}(\omega_{_{0}};\boldsymbol{c}) -  \boldsymbol{D}_{ss}^{^{(obs)}}(\omega_{_{0}})\right\Vert_{_{2}}^{^{2}} =\\
	 \sum_{ss \,=\,1}^{N_{_{ss}}}\, \left[\,\sum_{j^{^{(ss)}}\,=\,1}^{N_{_{Rx}}^{^{(ss)}} } \left| D^{^{(syn,enc)}}_{_{ss\,,j}}(\omega_{_{0}};\boldsymbol{c}_{_{k}})-D_{_{ss\,,j}}^{^{(obs)}}(\omega_{_{0}})\right|^{^{2}} \right]
						\end{split}
	\end{equation}
	where $\boldsymbol{D}_{ss}^{^{(obs)}}$ and $\boldsymbol{D}_{ss}^{^{(syn)}}$ are the stochastically phase-encoded representations of the observed and synthetic data respectively, in particular: 
		\begin{equation} \label{eq:eqDataRandom}
	\boldsymbol{D}_{ss}^{^{(obs)}}(\omega_{_{0}}) =  \sum_{s\, =\, 1}^{N_{_{ss}}}  \boldsymbol{d}^{^{(obs,\, enc)}}_{s}(\omega_{_{0}}) = \sum_{s\,=\, 1}^{N_{_{ss}}} a_{_{s}}  \, \boldsymbol{d}^{^{(obs)}}_{s}(\omega_{_{0}})
	\end{equation}
with an implicit dependence of the random phase $a_{_{s}}$ on the iteration number $k$ and on the super-source $ss$. Similarly to the case of $N_{_{s}}$ elementary sources, the encoded gradient is given by
		\begin{equation}
	\boldsymbol{\nabla_{_{c}}}^{^{(enc)}} \,=\,\mathfrak{Re}\left[\, \sum_{ss \,=\,1}^{N_{_{ss}}} \,\boldsymbol{\nabla_{_{c}}}^{^{{ss}}}\right]
	\end{equation}
	The encoded gradient, however, is not equivalent to the sum of the elementary source gradients, as it contains cross-talk terms in the form of iteration-dependent and super-source-dependent stochastic fluctuations
	\begin{equation}
	\boldsymbol{\nabla_{_{c}}} \,=\,\boldsymbol{\nabla_{_{c}}}^{^{(enc)}} + \text{stochastic cross-terms}
		\end{equation}
These can be averaged out by summing the contributions of multiple over-lapping super-sources  and/or re-withdrawing the random phase per super-source (multiple stochastic ensembles) at a fixed speed of sound iteration.

The source encoding approach described above is valid for any configuration of sources and receivers (as long as all the elementary sources making a super-source share the same set of receivers, which can be super-source-dependent), and it is equally valid both in 2D and in 3D. In the next section, we specialize the discussion to the geometry of interest.

	\subsection{Tomographic Geometry}
\begin{figure*}[htbp]
	\centering
	 \scalebox{0.53} 
	{\includegraphics{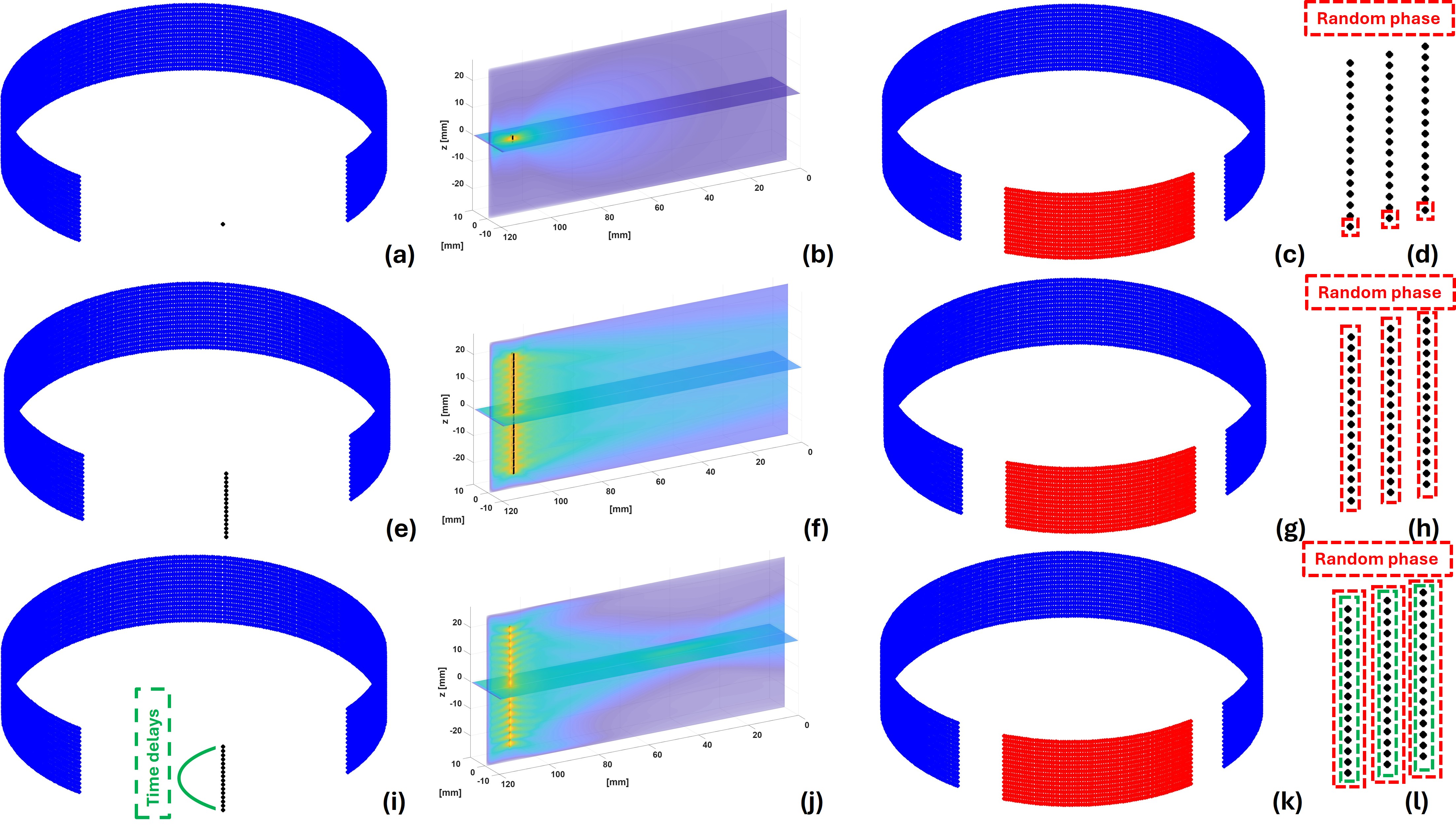}}
	 \caption{\textit{Configuration of super-sources and receivers.} (a) The elementary source, in black, corresponds to a quasi-omni-directional transmission (single array element), with the corresponding receivers in blue. (b) Elementary source field pattern (time-domain simulation, RF pulse centered at 700 kHz). (c) A super-shot, in red, with the corresponding receivers in blue. (d) Each elementary source in the super-shot (in this case, a single array element) is encoded with a different random phase (only a few are shown). The super-shot groups together $N_{_{Tx}}^{(ss)} = \frac{N}{6} \times N_{_{R}}$ elementary sources. (e) The elementary source, in black, corresponds to a line source, with the corresponding receivers in blue. (f) Elementary source field pattern. (g) A super-shot, in red, with the corresponding receivers in blue. (h) Each elementary source in the super-shot (in this case, a line source) is encoded with a different random phase. The super-shot groups together $N_{_{Tx}}^{(ss)} = \frac{N}{6}$ elementary sources. (i) The elementary source, in black, corresponds to a focused transmission, with the corresponding receivers in blue. (j) Elementary source field pattern. (k) A super-shot, in red, with the corresponding receivers in blue. (l) Each elementary source in the super-shot (in this case, a focused source) is encoded with a different random phase. The super-shot groups together $N_{_{Tx}}^{(ss)} = \frac{N}{6}$ elementary sources. [For visualization purposes, width and height of a single array element are displayed only in panels (b), (f) and (j).]}
	 \label{fig:fig1} 
	\end{figure*}

\begin{figure*}[htbp]
	\centering
	 \scalebox{0.53} 
	{\includegraphics{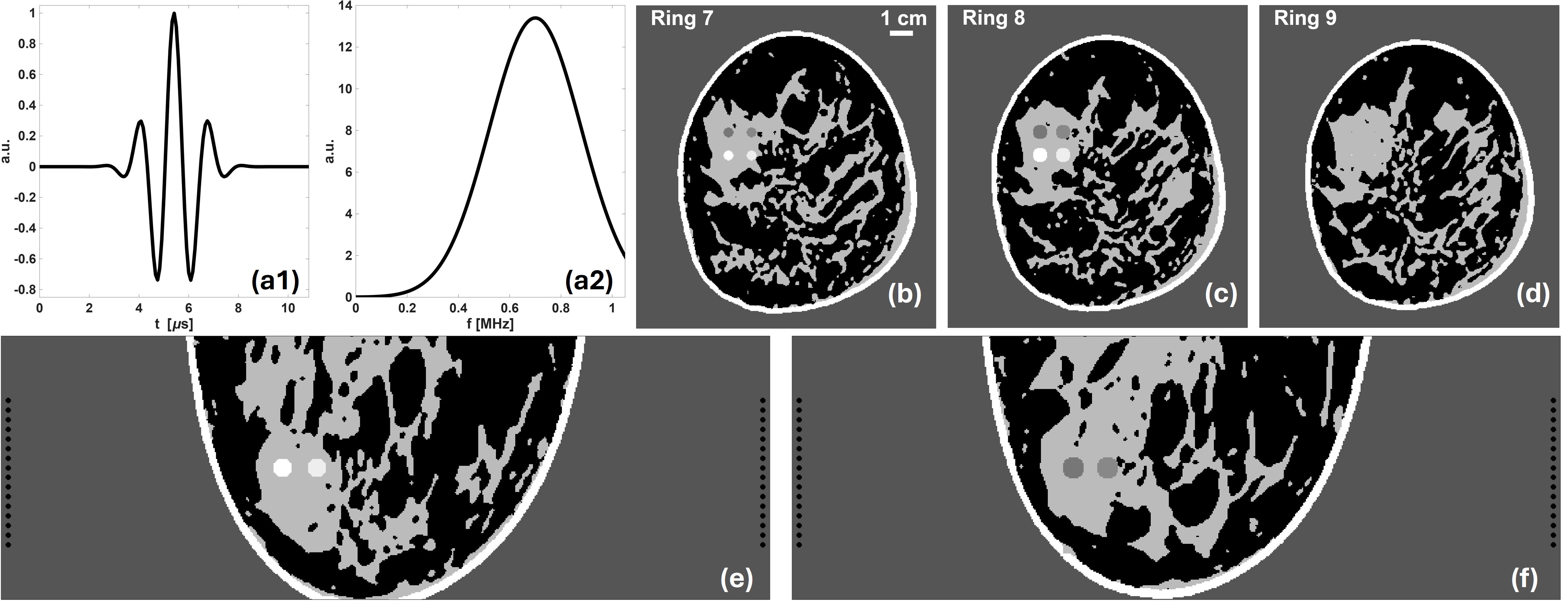}}
	 \caption{\textit{Excitation pulse and numerical phantom.} (a1) Excitation pulse, (a2) frequency spectrum. (b)-(d) Coronal sections in correspondence of the three rings co-located with the four inclusions. (e)-(f) Transverse sections in correspondence of the inclusions with a nominal positive and negative contrast respectively; in black, the location of the rings in elevation at two polar angles  (rings ordered from top to bottom). All speed of sound images are displayed on the same scale [1450,1600] m/s.}
	 \label{fig:fig2} 
	\end{figure*}

We focus the discussion around a cylindrical geometry. The \textit{raw} dataset, in the frequency domain, has a size of $N_{_{Rx}} \times N_{_{Tx}} \times N_{_{f}} \times N_{_{R}}^{^{(Rx)}} \times N_{_{R}}^{^{(Rx)}}$ complex samples, where $N_{_{Tx}}$ $\left(N_{_{Rx}}\right)$ is the number of transmitting (receiving) elements per ring, $N_{_{R}}^{^{(Tx)}}$ $\left(N_{_{R}}^{^{(Rx)}}\right)$ is the number of transmitting (receiving) rings and $N_{_{f}}$ is the number of frequencies to be inverted. We make the simplifying assumptions that $N_{_{Rx}} = N_{_{Tx}} = N$ and  $N_{_{R}}^{^{(Rx)}} = N_{_{R}}^{^{(Tx)}} = N_{_{R}}$. We refer to an array element as to a \textit{raw} source. We distinguish a few relevant cases depending on the nature of the elementary source:
\begin{itemize}
    \item \textit{Single array element (QODS)}. In this scenario, the elementary source and the raw source are equivalent, Fig. \ref{fig:fig1} (a)-(b). The super-shot is formed by grouping together $N_{_{Tx}}^{^{(ss)}}$ elementary sources \textit{per ring}; a random phase is applied to each elementary source, Fig. \ref{fig:fig1} (c)-(d), and to the corresponding \textit{raw} data-set (eq. (\ref{eq:eqDataRandom})).
    \item \textit{Line Source} (LS). In this scenario, a structured (\textit{coherent}) phase-encoding is first applied across individual raw sources at the same polar angle. The raw data-set is simply summed along the dimension corresponding to the transmitting rings, resulting into a (coherently-encoded) data-set of size $N \times N \times N_{_{f}} \times N_{_{R}}$ complex samples. In this case, the elementary source corresponds to a line source, Fig. \ref{fig:fig1} (e)-(f). The super-shot is formed by grouping together $N_{_{Tx}}^{^{(ss)}}$ elementary sources; a random phase is applied to each elementary source, Fig. \ref{fig:fig2} (g)-(h), and to the corresponding data-set \textit{on top} of the structured (coherent) phase-encoding.
      \item \textit{Focused Source} (FS). In this scenario, a structured (\textit{focusing}) phase-encoding is first applied across individual raw sources at the same polar angle. Time delays are calculated to achieve a nominal beam-width at a pre-selected focal depth (e.g. \cite{Cobbold}). The time delays are mapped into phase shifts. The raw data-set is phase-encoded accordingly along the dimension corresponding to the transmitting rings, resulting into a final data-set of size $N \times N \times N_{_{f}} \times N_{_{R}}$ complex samples. In this case, the elementary source corresponds to a focused  transmission (a beam), Fig. \ref{fig:fig1} (i)-(j). The super-shot is formed by grouping together $N_{_{Tx}}^{^{(ss)}}$ elementary sources; a random phase is applied to each elementary source, Fig. \ref{fig:fig1} (k)-(l), and to the corresponding raw data-set \textit{on top} of the structured (focusing) phase-encoding.

\end{itemize}

In all of the aforementioned cases, applying a random phase to the elementary source aims to preserve its field pattern and its associated resolution, which would otherwise be destroyed by a structured (e.g. geometric or synthetic) phase encoding. Due to the symmetry of the geometry under consideration, the super-sources are evenly spaced along the cylindrical surface; the receivers rotate accordingly. It's worth stressing that the receiving elements are quasi-omni-directional in all of the considered cases. Apart from the conceptual difference when dealing with the three different elementary sources, there is, however, a significant difference in the computational gains when comparing against the conventional FD-FWI formulation. In the latter, one has to solve $2 \times N \times N_{_{_R}}$  ($s=$ QODS) and $2 \times N$ ($s=$ LS or FS) Helmholtz equations per frequency iteration respectively in order to compute the gradient. In the formulation with stochastic phase-encoding, one has to solve $2 \times N_{_{ss}}$ Helmholtz equations in total, regardless of the nature of the elementary source. For the geometry under consideration, the computational gain is large in all the three cases, with a dramatic improvement when the elementary source is a single array element. Of course, image quality has to be compared across the three different scenarios in relation to a specific clinical task (e.g. high sensitivity). This is investigated in Section \ref{Results}.

\begin{table}
		\centering
				\caption{Parameters for forward simulations.}
		\begin{tabular}{|p{100pt}|p{100pt}|}
		\hline
		Grid Size & 600 x 600 x 140  \\
         Voxel & 0.4 mm \\
		Time Step & 83.3 ns (12 MHz) \\
		Time Samples & 2400 \\
		Frequency & 700 kHz (central, pulsed) \\
		Radius &  110 mm \\
         Ring-Array Elements $N$ & 512 \\
		Number of Rings $N_{_{R}}$ & 16 \\
		Sampling in elevation & 3 mm \\
	    	Element Width &  0.7 mm \\
	    	Element Height & 2 mm \\
		\hline
		\end{tabular}
		\label{Tab:Tab1}
		\end{table}	
		
					\begin{table}
					\centering 
		\caption{Parameters for frequency-domain reconstruction.}
		\label{Tab:Tab2}
		 \begin{tabular}{|p{80pt}|p{60pt}|p{40pt}|p{25pt}|}
		\hline
		 Inverted Frequencies & Grid Size & Voxel (mm) & $N^{^{\text{(Born)}}}$ \\
		 (100 kHz step) &  & & \\
								\hline		
			300 - 700   & 480 x 480 x 120 &  0.5   & 200\\
									800 - 1000   & 600 x 600 x 150 &  0.4 & 300\\
						\hline
		\multicolumn{3}{p{180pt}}{}\\
				\end{tabular}
						 \begin{tabular}{|p{110pt}|p{120pt}|}
		\hline
		Super-shots configuration (3D) & Parameters\\
		\hline
		 Number of super-shots  & $N_{_{ss}} = 64$ \\
	     Number of elementary sources & $N_{_{Tx}}^{(ss)} = \frac{N}{6} \times N_{_{R}}$ ($s=$ QODS) \\
	       & $N_{_{Tx}}^{(ss)} = \frac{N}{6}$  ($s=$ LS or FS) \\
		 Number of receiving elements  & $N_{_{Rx}}^{(ss)}=\frac{3N}{4} \times N_{_{R}}$ \\
		 Number of stochastic ensembles &  $N_{_{PE}} = 1$\\
		 		\hline
		Sources configuration (2D) & Parameters\\
				\hline
		Number of sources  & $N_{_{s}} = 512$ \\
				Number of receiving elements  & $N_{_{Rx}}^{^{(s)}} = \frac{3N}{4} $ \\
		\hline
			\multicolumn{2}{p{230pt}}{}\\
		\end{tabular}
		\end{table}

	 \begin{figure*}[htbp]
	\centering
	 \scalebox{0.17} 
	{\includegraphics{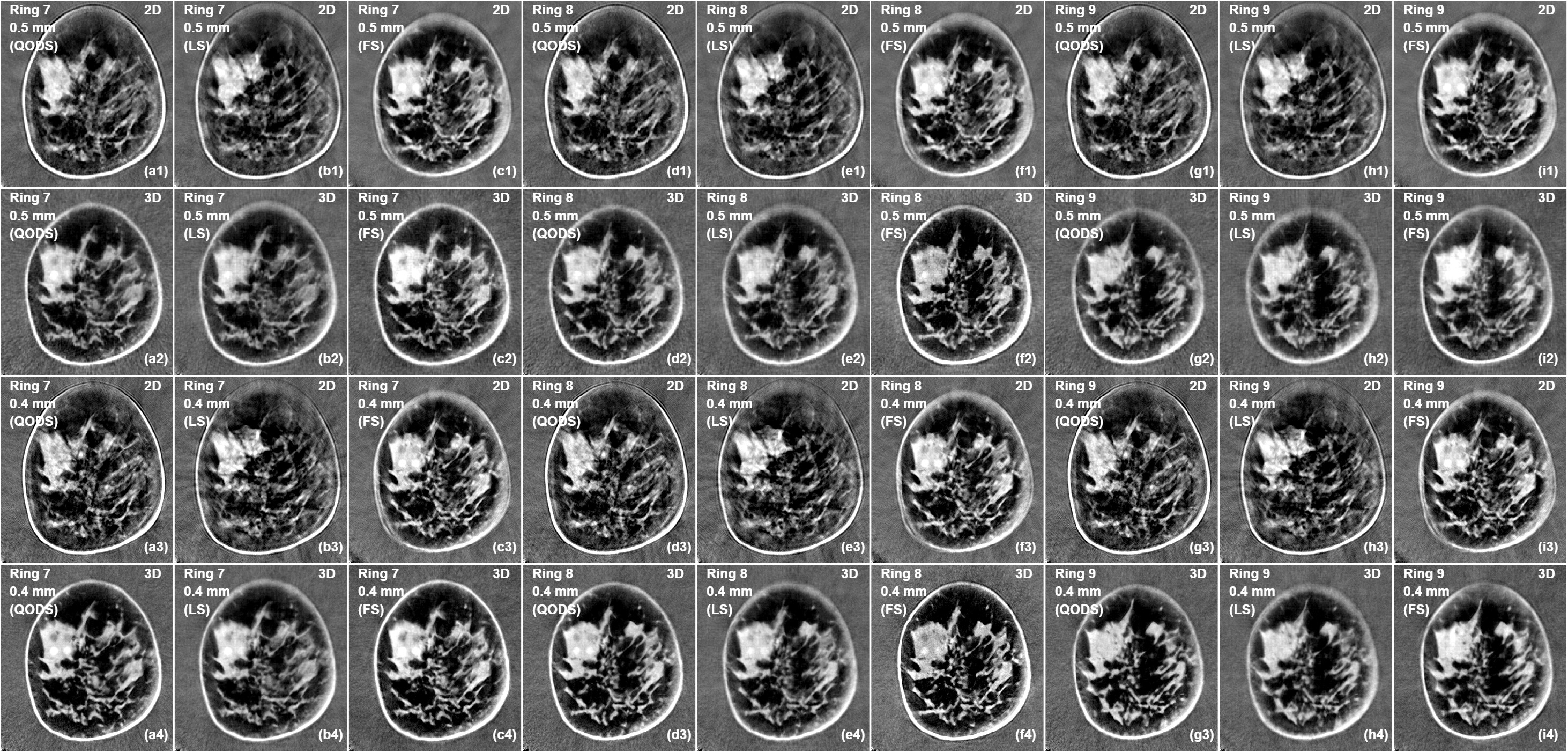}}
	 \caption{\textit{Coronal sections (inclusions). Comparison between 3D reconstruction and 2D reconstruction.} (a1) 2D reconstruction when ring 7 is transmitting and receiving, in the bandwidth 300-700 kHz (0.5 mm). (b1) 2D reconstruction when ring 7 is receiving (line source transmission),  in the bandwidth 300-700 kHz (0.5 mm). (c1) 2D reconstruction when ring 7 is receiving (focused transmission), in the bandwidth 300-700 kHz (0.5 mm). (a2) 3D reconstruction (quasi-omni-directional transmission), in the bandwidth 300-700 kHz (0.5 mm). (b2) 3D reconstruction (line source transmission), in the bandwidth 300-700 kHz (0.5 mm). (c2) 3D reconstruction (focused transmission), in the bandwidth 300-700 kHz (0.5 mm). (a3) 2D reconstruction when ring 7 is transmitting and receiving, in the bandwidth 800-1000 kHz (0.4 mm). (b3) 2D reconstruction when ring 7 is receiving (line source transmission), in the bandwidth 800-1000 kHz (0.4 mm). (c3) 2D reconstruction when ring 7 is receiving (focused transmission), in the bandwidth 800-1000 kHz (0.4 mm). (a4) 3D reconstruction (quasi-omni-directional transmission), in the bandwidth 800-1000 kHz (0.4 mm). (b4) 3D reconstruction (line source transmission), in the bandwidth 800-1000 kHz (0.4 mm). (c4) 3D reconstruction (focused transmission), in the bandwidth 800-1000 kHz (0.4 mm). (d)-(f), (g)-(i) same as (a)-(c) for ring 8 and ring 9 respectively. All images are displayed on the same scale [1450,1600] m/s.}
	 \label{fig:fig3} 
	\end{figure*}
		 \begin{figure*}[htbp]
	\centering
	 \scalebox{0.22} 
	{\includegraphics{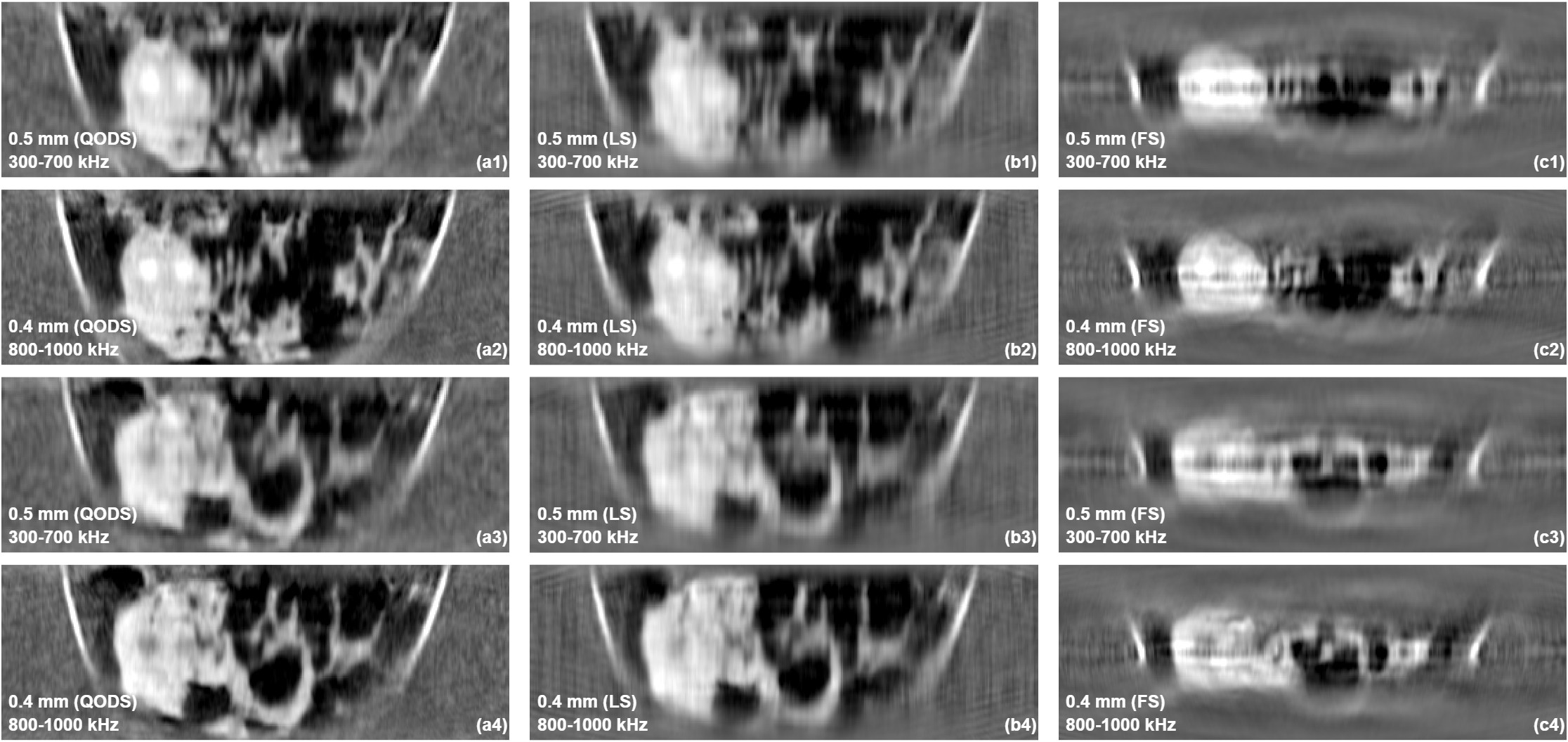}}
	 \caption{\textit{Transverse Sections (inclusions). 3D reconstruction.} (a1)-(a2) 3D reconstruction (quasi-omni-directional transmission) in the bandwidths 300-700 kHz (0.5  mm) and 800-1000 kHz (0.4 mm): sections in correspondence of the inclusions with a positive contrast. (a3)-(a4) 3D reconstruction (quasi-omni-directional transmission) in the bandwidths 300-700 kHz (0.5  mm) and 800-1000 kHz (0.4 mm): sections in correspondence of the inclusions with a negative contrast. (b1)-(b4) and (c1)-(c4) same as (a1)-(a4) for a line source and a focused transmission respectively.}
	 \label{fig:fig4} 
	\end{figure*}
	 \begin{figure*}[htbp]
	\centering
	 \scalebox{0.22} 
	{\includegraphics{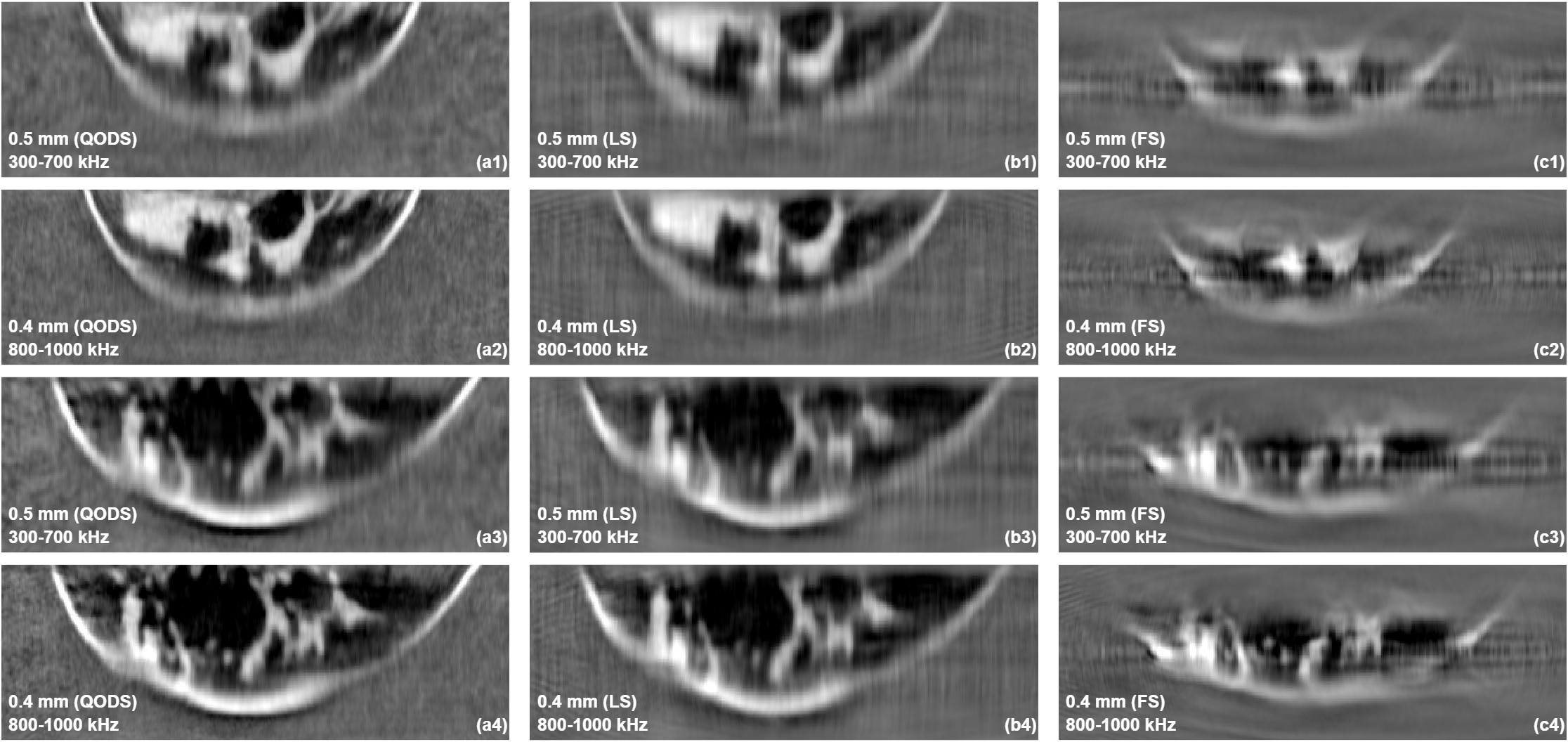}}
	 \caption{\textit{Transverse Sections (nipple area). 3D reconstruction.} (a1)-(a2) 3D reconstruction (quasi-omni-directional transmission) in the bandwidths 300-700 kHz (0.5  mm) and 800-1000 kHz (0.4 mm): sections in correspondence of $y = 0$ mm. (a3)-(a4) 3D reconstruction (quasi-omni-directional transmission) in the bandwidths 300-700 kHz (0.5  mm) and 800-1000 kHz (0.4 mm): sections in correspondence of $x = 0$ mm. (b1)-(b4) and (c1)-(c4) same as (a1)-(a4) for a line source and for a focused transmission respectively.}
	 \label{fig:fig5} 
	\end{figure*}
	 \begin{figure*}[htbp]
	\centering
	 \scalebox{0.18} 
	{\includegraphics{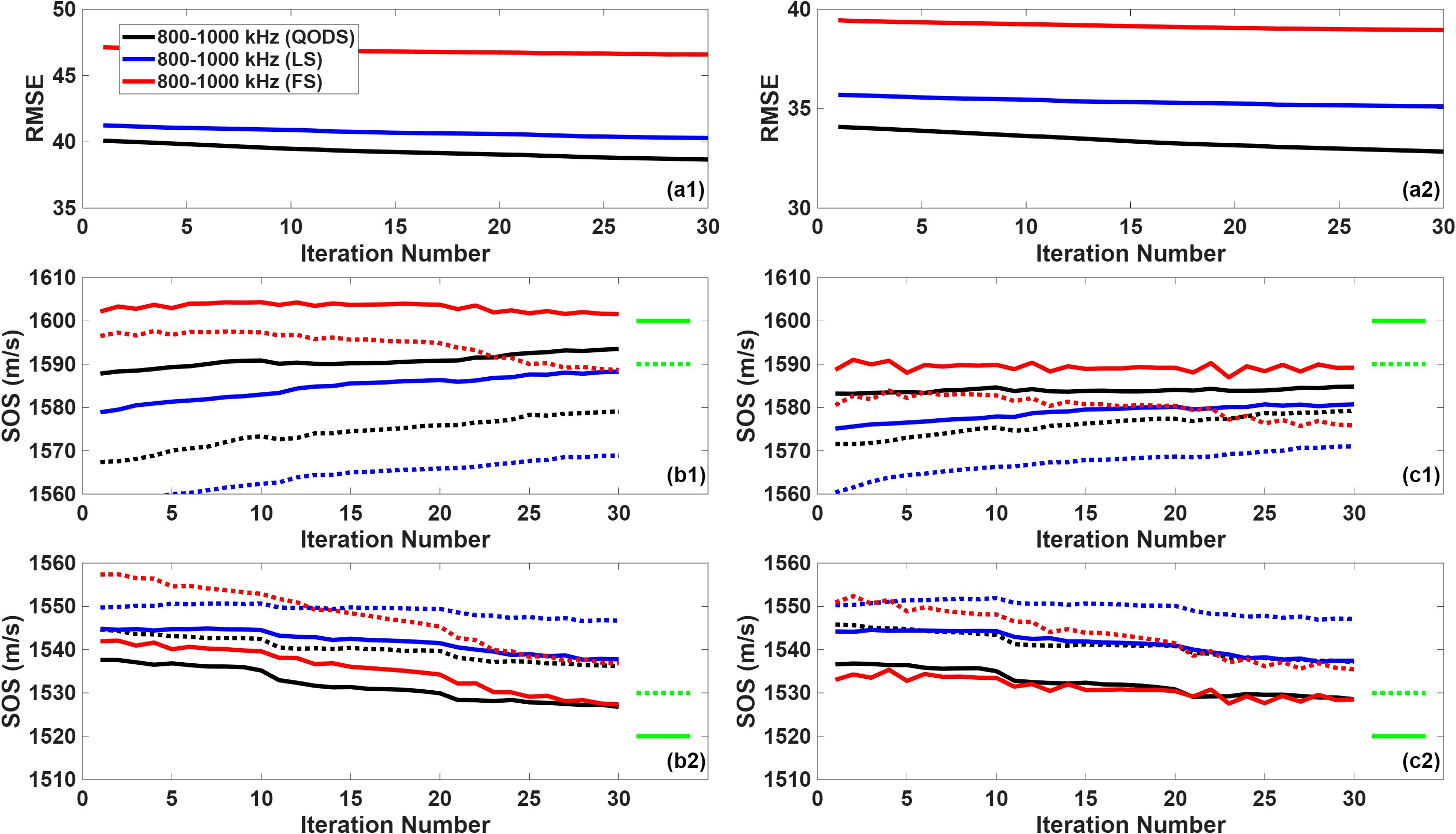}}
	 \caption{\textit{Quantitative assessment of image quality in the 800-1000 kHz bandwidth. 3D reconstruction.} (a1)-(a2) Volume RMSE for the reconstructed images shown in Figs. \ref{fig:fig4} and \ref{fig:fig5} respectively. (b1) Reconstructed speed of sound values for the spherical inclusions with positive contrast (coronal sections, ring 7). Continuous/dotted lines correspond to the inclusion with a larger/smaller contrast. True values in green. (b2) Reconstructed speed of sound values for the spherical inclusions with a negative contrast (coronal sections, ring 7). (c1) Reconstructed speed of sound values for the spherical inclusions with a positive contrast (transverse sections). (c2) Reconstructed speed of sound values for the spherical inclusions with a negative contrast (transverse sections).}
	 \label{fig:fig6} 
	\end{figure*}

	\section{Results} \label{Results}
To keep the computing times reasonable and preserve the information contained in the forward data, we have considered a cylindrical geometry with $N_{_{R}} = 16$ rings (3 mm sampling in elevation), each with $N = N_{_{Tx}} = N_{_{Rx}} = 512$ array elements; single-element width and height have also been taken into account (0.7 mm $\times$ 2 mm). This results into solving $N \times N_{_{R}}$ wave equations in time-domain on a 3D grid of size 240 mm $\times$ 240 mm  $\times$ 56 mm and voxel 0.4 mm. The computing times for the forward data-set are approximately 3 weeks (kWave on a single 16 GB GPU card, NVIDIA RTX A4000); the storage amounts to 600 GB. To meet the stability requirements of the wave solver, the RF pulse has been centered around 700 kHz, Fig. \ref{fig:fig2} (a1)-(a2). The parameters  for the forward simulations are summarized in Table \ref{Tab:Tab1}. When the elementary source is a single element, frequency domain samples are obtained via a standard FFT. When the elementary source is a line source, the frequency samples are summed along the dimension corresponding to the transmitting rings. When the elementary source is a focused one, the frequency samples are summed along the dimension corresponding to the transmitting rings after applying focusing phase shifts; the latter assume a nominal focal depth of 10 cm.

The numerical phantom is described in \cite{Anastasio4}, and it is representative of a dense breast with fatty tissue (1450 m/s), skin layer (1700 m/s) and significant presence of glandular tissue (1560 m/s). To test the reconstruction capabilities of the proposed approach, 4 spherical inclusions of 6 mm diameter and of varying positive and negative acoustic contrast ($\pm 30$ m/s, $\pm 40$ m/s) have been added to the glandular tissue. The inclusions sit between rings 7 and 9, in the central part of the portion of the breast under consideration, Fig. \ref{fig:fig2} (b)-(d). The chosen geometry allows to insonify a portion of an anatomical breast that extends roughly for 5 cm in elevation, Fig. \ref{fig:fig2} (e)-(f). A second data-set is obtained by translating vertically the 16 rings by 2.5 cm to image the region of the breast in proximity of the nipple.

The reconstruction algorithm is run as described in \cite{ForteIEEE2026}. Wherever possible, to avoid inverse crime, the reconstruction grid voxel is larger than the forward one. To introduce further mismatch with respect to the forward model, array elements have been modeled as perfectly omni-directional in the reconstruction phase. The nominal coordinates of (the centroid of) the ring-array elements have been mapped to a single grid point in the reconstruction grid. The Helmholtz equation has been solved via the modified Born series \cite{Ivo}, retaining 200 or 300 terms in the Born series. The modified version of the Born-Neumann series is not necessarily needed in breast imaging. It has been employed because this wave solver has been shown to be successful in the inversion problem in 2D and 3D in \cite{Duric3D} and \cite{ForteIEEE2026}, and in preparation for future applications to body parts that may have a much stronger acoustic contrast than a human breast or soft tissue \cite{Wiskin3}-\cite{Wiskin4}. The code for the wave solver and the reconstruction algorithm has been written by the author and implemented in MATLAB; in particular, forward and adjoint equation per super-source are solved sequentially. For comparison, 2D reconstructions are also run on a 2D grid of same size and same pixel with standard FD-FWI as described in \cite{ForteIEEE2026}, with the LU-based wave solver available from \cite{DuricLast}. The parameters for image reconstruction are summarized in Table \ref{Tab:Tab2}.

\subsection{Coronal sections}
We start by visually assessing image quality for coronal sections and compare against a standard 2D reconstruction. We limit the discussion of the coronal sections to the ones in correspondence of the 4 inclusions. 

\subsubsection{Reconstruction below 700 kHz}
In the following, all the 2D and 3D reconstructions up to 700 kHz assume a flat initial guess (1500 m/s).  The coronal sections in correspondence of the rings 7-9, and for the three different elementary sources described above, are shown in Fig. \ref{fig:fig3} (a1)-(b1)-...-(i1) for the 2D reconstruction and in Fig. \ref{fig:fig3} (a2)-(b2)-...-(i2) for the 3D reconstruction. It is possible to observe a strong presence of out-of-plane scattering artifacts in all of the 2D reconstructions: each section is crowded with anatomical features not present in the true object (both in proximity of the skin layer and in the interior region), whereas the 3D ones show a clean reconstruction, more representative of the true object. Also, the 2D reconstructions significantly overestimate the size of the corresponding breast slice: this is less true for a focused transmission due to its beam-width. As expected from the field pattern of the corresponding elementary source, the 2D reconstructions are problematic when it comes to detect the four inclusions, with the exception of a focused transmission: in the latter, however, the 4 inclusions, albeit detected, are incorrectly displayed on ring 9, fact due to the out-of-plane scattering being projected onto the imaging planes in the beam-width of the focused transmission. The 3D reconstructions show the presence of the two inclusions with the larger contrast (in absolute value) across rings 7-8, and only show partial shadows of the two inclusions with a smaller contrast.

For 64 super-sources, the total computing times are around 14 hours on a desktop pc (5 inverted frequencies, 10 iterations per frequency, 0.5 mm voxel).

\subsubsection{Reconstruction above 700 kHz}
In the following, the initial velocity model for the 2D and 3D reconstructions in the 800 kHz - 1 MHz bandwidth has been obtained by interpolating the corresponding final estimate for the speed of sound up to 700 kHz (0.5 mm) onto a finer grid (0.4 mm). Results are shown in Fig. \ref{fig:fig3}. The 2D reconstructions still struggle to detect the four inclusions, with the exception of a focused transmission, which, however, misplaces the correct locations. With subtle variability in image quality, all the 3D reconstructions are capable to detect the four inclusions and place them correctly across rings 7-8 with an image contrast in agreement with the true acoustic contrast. In terms of perceived image quality, the sections in correspondence of ring 8 show a deteriorated image contrast when the elementary source is a line one, Fig. \ref{fig:fig3} (e4), or a focused one, Fig. \ref{fig:fig3} (f4), with respect to the quasi-omni-directional source, Fig. \ref{fig:fig3} (d4). 

We have to note that some inverse crime is still committed in this instance, as the reconstruction grid and the forward grid are essentially the same. Also, the frequencies in the 800-1000 kHz bandwidth are supported between 5 and 3 ppw in the forward modeling.

For 64 super-sources, the total computing times are around 24 hours on a desktop pc (3 inverted frequencies, 10 iterations per frequency, 0.4 mm voxel).

\subsection{Transverse sections}
Transverse sections in correspondence of the inclusions with a positive contrast, and for the three different elementary sources described above, are shown in Fig. \ref{fig:fig4} (a1)-(a2), (b1)-(b2), (c1)-(c2); the ones in correspondence of the inclusions with a negative contrast are shown in Fig. \ref{fig:fig4} (a3)-(a4), (b3)-(b4), (c3)-(c3). The most immediate result is that the focused transmission is capable to reconstruct only a partial volume, in agreement with the field pattern in elevation of the elementary source; artifacts in correspondence of the beam are also visible. A visual analysis of the two other transmissions suggests that the line source one is more blurred in elevation (in agreement with the field pattern in elevation of the elementary source), which results into reduced image contrast and tissue separation in detecting all the four inclusions. The latter are more clearly delineated when the elementary source is a quasi-omni-directional one.

For the nipple area, transverse sections in correspondence of $x = 0$ mm and $y = 0$ mm are shown in Fig. \ref{fig:fig5}; these show similar qualitative features. One can further observe that the skin layer which is not well delineated in the bottom parts of the images in Fig. \ref{fig:fig4} is more pronounced in Fig. \ref{fig:fig5}, especially when the elementary source is quasi-omni-directional. Further manifestation of the phenomenon of the skin layer erosion is also visible in all cases and it is due to the combination of the breast curvature and cylindrical geometry.

\subsection{Quantitative Assessment of Image Quality} 
Some metrics for a quantitative assessment of image quality are shown in Fig. \ref{fig:fig6}; we limit the discussion to the 800-1000 kHz bandwidth. These include the volume RMSE, Fig. \ref{fig:fig6} (a1)-(a2), for the images shown in Figs. \ref{fig:fig4} and \ref{fig:fig5} respectively: the decaying profiles are in agreement with a subjective assessment of image quality, i.e. with the interpretation that a quasi-omni-directional transmission provides a more complete reconstruction of the insonified volume. Reconstructed speed of sound values for the 4 spherical inclusions are shown in Fig. \ref{fig:fig6} (b1)-(b2) for a coronal section co-located with Ring 7 and in Fig. \ref{fig:fig6} (c1)-(c2) for the corresponding transverse sections; values have been averaged over a manually segmented region overlapping with the interior of each inclusion. Overall, the reconstructed values exhibit profiles which, across iterations, move away from the background tissue (1560 m/s) and tend to converge towards the true values. In all cases, the final image contrast is representative of the true underlying speed of sound contrast (among the inclusions and with respect to the background tissue); for each elementary source, reconstructed speed of sound values between the two positive/negative contrast inclusions tend to differ by 10-15 m/s, in very good agreement with the nominal value. The reconstructed values show slightly better agreement on the coronal section, perhaps unsurprisingly as the chosen geometry only partially encloses the insonified region.

Finally, all of the cost functions for the 2D and 3D reconstructions exhibit similar decaying profiles to the ones shown in \cite{ForteIEEE2026} and they are not included for the sake of brevity.

\section{Discussion and Conclusions} \label{sec:end}

By employing a limited number of randomly phase-encoded elementary sources in combination with FD-FWI one can accurately reconstruct a portion of an anatomical breast in 3D. This is possible regardless of the nature of the elementary source (quasi-omni-directional, line or focused): this is expected and ultimately due to the linearity properties of wave propagation with respect to the source term. The reconstructed images prove that it is possible to detect mm-size masses in breast tissue with an acoustic contrast as low as 2$\%$ with respect to glandular tissue, thus opening the concrete possibility to the design and the deployment of medical devices with \textit{high} sensitivity. These results add up to the evidence provided with standard 3D FD-FWI in \cite{Duric3D}, which has considered larger masses and with a larger acoustic contrast.

Guidelines for breast MRI imaging \cite{Kinkel} recommend an in-plane resolution of 1 mm and a transverse one not larger than 2.5 mm to guarantee detection and characterization of masses larger than or equal to 5 mm. The rationale for these recommendations can be justified by the inherent high image contrast in a T1-weighted protocol due to the injection of a positive contrast agent. Acoustic contrast is natively low, so the requirement to have an isotropic voxel below 0.4 mm may be crucial\footnote{Although computing times obviously increase, memory requirements for Born-series approaches are not demanding; non-isotropic voxels can also be dealt with.}. Further improvement in image quality (e.g. resolution and contrast) can undoubtedly be achieved by inverting on a smaller voxel (0.3 mm) and including the directionality of the array elements in the inversion phase. Testing this hypothesis on synthetic data requires a forward grid with a voxel around 0.2 mm and an RF pulse centered at 1 MHz for all the inverted frequencies to be supported with enough ppw in the inversion phase, as in \cite{ForteIEEE2026}. Initial tests suggest that the computing times for the same cylindrical geometry and \textit{one} imaging volume may require up to a year on multiple (large memory)  GPUs, hence these investigations are left to the \textit{design} phase of the device. The storage requirements for a few imaging volumes quickly exceed a few TB, which also raises concerns about the practicality of processing time-domain data in a clinical setting for tasks other than a DTFT.  

The approach is transferable to other tomographic geometries topologically (or not topologically) equivalent to a cylindrical one, e.g. an imaging hemi-sphere, or a box, 4 or more matrix probes. In the latter case, production lines may already satisfy the requirements for tomographic imaging. 

The computing times for all of the above configurations are the same, as we have fixed the number of super-sources $N_{_{ss}} = 64$. With a professional implementation of the reconstruction algorithm in CUDA C/C++, the computing times can be reduced by a factor of 4, and by parallelizing 8 super-sources on 8 low memory or 4 large memory GPUs, the reconstruction of a breast volume extending for 5 cm can be achieved within one hour. The computing times become then feasible for medical imaging applications and do allow for a rigorous investigation of whether image quality may or may not satisfy met and unmet clinical needs (e.g. high sensitivity).

In terms of acquisition times for a real device, if tissue penetration requirements are met for all of the elementary sources, a line or a focused transmission is of course faster, with a gain of $N_{_{R}}$ with respect to the case of a single element-based transmission. For the parameters above, the nominal acquisition times per imaging volume are 26.21 s ($s$ = single element, $N_{_{Tx}} \times N_{_{R}}\times N_{_{R}}$ transmission events) and 1.64 s ($s$ = line source or focused transmission, $N_{_{Tx}} \times N_{_{R}}$ transmission events), assuming that $N_{_{Rx}} = 512$ elements receive in parallel (i.e. 512 independent ADC channels sampling at RF). The results above suggest that it is possible to translate an elevational aperture of 5 cm by 2-3 cm to achieve a reasonably accurate reconstruction of an anatomical breast (around 7 cm in elevation) with a single-element-based-transmission, or  with a line-source-based-transmission with inferior image quality, whereas a focused transmission requires a finer sampling in elevation, thus further extending the acquisition times and the storage burden over the other two configurations\footnote{On the other hand, if carefully designed, the focus of the transmission and the beam-width may be \textit{dynamic}, i.e. they can be adjusted to follow the breast curvature. Issues about spatial sampling and grating lobes have to be addressed of course. For comparison, the system described in \cite{Wiskin2}-\cite{Wiskin4} has a fixed focus at 6 cm (lens), it collects data on an elevational aperture of 2 cm and it scans the vertical dimension with a sampling of 2 mm.}. Even when considering the need to stretch the PRF to prevent ambiguous echoes from earlier transmissions, the actual acquisition times are within few minutes per breast, and thus acceptable for diagnostic imaging.

Transmitting up to 100 V${_{_{\text{pp}}}}$ from single array elements, applying time delays for beam formation, and receiving from 512 or more channels in parallel, are already within the capabilities of modern medical ultrasound scanners. One or two GPUs are often integrated; the possibility of streaming frequency samples to a remote server with multiple GPUs is also a concrete one. Thus, in conclusions, except for the need of translating the device along the breast (feature that can be eliminated by designing a modular system), it is in principle possible to integrate a system similar to the one described in this manuscript into modern and next generation clinical ultrasound consoles.

	 \printbibliography

	\end{document}